\documentclass[pra,aps,twocolumn,floatfix,showpacs]{revtex4}
\usepackage{graphicx,amsmath,amssymb,times}

\begin{document}
\title{Mean-field theory for the Mott insulator--paired superfluid transition in the two-species Bose-Hubbard model}
\author{M. Iskin}
\affiliation{Department of Physics, Ko\c c University, Rumelifeneri Yolu, 34450 Sariyer, Istanbul, Turkey}
\date{\today}

\begin{abstract}
The standard mean-field theory for the Mott insulator--superfluid phase transition 
is not sufficient to describe the Mott insulator--paired superfluid phase transition. 
Therefore, by restricting the two-species Bose-Hubbard Hamiltonian to the subspace 
of paired particles, and using perturbation theory, here we derive an analytic 
mean-field expression for the Mott insulator--paired superfluid transition boundary.
\end{abstract}

\maketitle

\section{Introduction}
\label{sec:introduction}

Following the recent observation of Mott insulator--superfluid transition 
with ultracold atomic Bose gases loaded into optical 
lattices~\cite{greiner02, stoferle04, spielman07, spielman08}, there has 
been an intense theoretical activity in analyzing many Hubbard-type 
lattice models~\cite{bloch08}. Among them the two-species Bose-Hubbard 
model, which can be studied with two-component Bose gases loaded into 
optical lattices, is one of the most popular. This is because,
in addition to the Mott insulator and single-species
superfluid phases, it has been predicted that this model has at least 
two additional phases: an incompressible super-counter flow and a 
compressible paired superfluid 
phase~\cite{kuklov03, altman03, kuklov04, isacsson05, arguelles07, trefzger09}. 

Our main interest here is in the latter phase, where a direct 
transition from the Mott insulator to the paired superfluid phase 
(superfluidity of composite bosons, i.e. Bose-Bose pairs) has been
predicted, when both species have integer fillings and the 
interspecies interaction is sufficiently large and attractive.
In this paper, we derive an analytic mean-field expression for the 
Mott insulator--paired superfluid transition boundary in the two-species
Bose-Hubbard model. The remaining paper is organized as follows. 
After introducing the model Hamiltonian in Sec.~\ref{sec:tsbh}, first 
we derived the mean-field theory in Sec.~\ref{sec:mft}, and then presented 
typical phase diagrams in Sec.~\ref{sec:pd}. A brief summary of our
conclusions is given in Sec.~\ref{sec:conclusions}.

\section{Two-species Bose-Hubbard Model}
\label{sec:tsbh}

The two-species Bose-Hubbard Hamiltonian is given by,
\begin{align}
\label{eqn:ebhh}
H = &- \sum_{i,j, \sigma} t_{ij,\sigma} b_{i,\sigma}^\dagger b_{j,\sigma} 
  + \sum_{i,\sigma} \frac{U_{\sigma\sigma}}{2} \widehat{n}_{i,\sigma} (\widehat{n}_{i,\sigma}-1) \nonumber \\
&+ U_{\uparrow\downarrow}\sum_{i} \widehat{n}_{i,\uparrow} \widehat{n}_{i,\downarrow} - \sum_{i,\sigma} \mu_\sigma \widehat{n}_{i,\sigma},
\end{align}
where the pseudo-spin $\sigma \equiv \{\uparrow, \downarrow\}$ labels the 
trapped hyperfine states of a given species of bosons, or labels different 
types of bosons in a two-species mixture, $t_{ij, \sigma}$ is the 
tunneling (or hopping) matrix between sites $i$ and $j$, 
$b_{i,\sigma}^\dagger$ ($b_{i,\sigma}$) is the boson creation (annihilation) 
and $\widehat{n}_{i,\sigma} = b_{i,\sigma}^\dagger b_{i,\sigma}$ 
is the boson number operator at site $i$, $U_{\sigma\sigma'}$ is the strength 
of the onsite boson-boson interaction between $\sigma$ and $\sigma'$ 
components, and $\mu_\sigma$ is the chemical potential.
In this manuscript, we consider a $d$-dimensional hypercubic lattice,
for which we assume $t_{ij,\sigma}$ is a real symmetric matrix with elements
$t_{ij,\sigma} = t_\sigma \ge 0$ for $i$ and $j$ nearest neighbors and $0$ otherwise.
We take the intraspecies interactions to be repulsive ($\{U_{\uparrow\uparrow}, 
U_{\downarrow\downarrow}\} > 0$) and the interspecies interaction to be attractive 
($U_{\uparrow\downarrow} < 0)$ such that 
$U_{\uparrow\uparrow} U_{\downarrow\downarrow} > U_{\uparrow\downarrow}^2$,
to guarantee the stability of the mixture against collapse. 

For sufficiently attractive $U_{\uparrow\downarrow}$, it is well-established 
that~\cite{kuklov03, altman03, kuklov04, isacsson05, arguelles07, trefzger09} 
instead of a direct transition from the Mott insulator to a single particle 
superfluid phase, the transition is from the Mott insulator to a paired 
superfluid phase (superfluidity of composite bosons, i.e. Bose-Bose pairs). 
In fact, in the limit when $\{t_\uparrow, t_\downarrow\} \to 0$, 
it can be shown that the transition is from the Mott insulator to a 
paired superfluid phase for all $U_{\uparrow\downarrow} < 0$~\cite{iskin10}.

\subsection{Mean-field theory}
\label{sec:mft}

In the single-species Bose-Hubbard model, the standard mean-field theory, 
where the boson creation and annihilation operators are approximated
by their expectation values, e.g. 
$
b_{i,\sigma} = \langle b_{i,\sigma} \rangle + \delta b_{i,\sigma},
$
has proved to be very useful in understanding the qualitative features 
of the Mott insulator--single species superfluid phase transition~\cite{bloch08}. 
This is simply because the transition is driven by the first-order hopping 
effects. However, the Mott insulator--paired superfluid transition is 
driven by the second-order hopping effects, and therefore, the standard 
mean-field theory is not sufficient. This difficulty could be overcome 
by restricting the Hamiltonian to the subspace of paired particles, and 
including the second-order hopping effects through second-order 
perturbation theory~\cite{kuklov03, kuklov04, altman03, trefzger09}. 

For the Hamiltonian given in Eq.~(\ref{eqn:ebhh}), we have recently 
calculated the two-particle and two-hole excitation energies 
(i.e. energy costs for adding and removing two-particles, respectively) 
up to third order in the hoppings. Assuming 
$
\{ U_{\sigma\sigma}, |U_{\uparrow\downarrow}|, 2U_{\sigma\sigma}+U_{\uparrow\downarrow} \} \gg t_\sigma,
$
the two-particle excitation energy was found to be~\cite{iskin10}
\begin{align}
\label{eqn:en-tpar}
E_{\rm p} &= U_{\uparrow\downarrow} (n_\uparrow+n_\downarrow+1)
+ \sum_\sigma \left( U_{\sigma\sigma} n_\sigma - \mu_\sigma \right) \nonumber \\
&+ \sum_\sigma
	\left[ \frac{(n_\sigma+1)^2}{U_{\uparrow\downarrow}}
	- \frac{n_\sigma(n_\sigma+2)}{2U_{\sigma\sigma}+U_{\uparrow\downarrow}} 
	+ \frac{2n_\sigma(n_\sigma+1)}{U_{\sigma\sigma}} 
	\right] z t_\sigma^2 \nonumber \\
&+ \frac{2(n_\uparrow+1)(n_\downarrow+1)}{U_{\uparrow\downarrow}} z t_\uparrow t_\downarrow,
\end{align}
where $z = 2d$ is the coordination number.
Similarly, the two-hole excitation energy was found to be~\cite{iskin10}
\begin{align}
\label{eqn:en-thol}
E_{\rm h} &= - U_{\uparrow\downarrow} (n_\uparrow+n_\downarrow-1)
- \sum_\sigma \left[ U_{\sigma\sigma} (n_\sigma-1) - \mu_\sigma \right] \nonumber \\
&+ \sum_\sigma
	\left[ \frac{n_\sigma^2}{U_{\uparrow\downarrow}} 
	- \frac{(n_\sigma^2-1)}{2U_{\sigma\sigma}+U_{\uparrow\downarrow}}
	+ \frac{2n_\sigma(n_\sigma+1)}{U_{\sigma\sigma}} 
	\right] z t_\sigma^2 \nonumber \\
&+ \frac{2n_\uparrow n_\downarrow}{U_{\uparrow\downarrow}} z t_\uparrow t_\downarrow,
\end{align}
The accuracy of Eqs.~(\ref{eqn:en-tpar}) and~(\ref{eqn:en-thol}) are checked
via exact small-cluster (two-site) calculations. In addition, in the limit when 
$t_\uparrow = t_\downarrow = t$, $U_{\uparrow\uparrow} = U_{\downarrow\downarrow} = U_0$, 
$U_{\uparrow\downarrow} = U'$, $n_\uparrow = n_\downarrow = n_0$, 
$\mu_\uparrow = \mu_\downarrow = \mu$, and $z = 2$ (or $d = 1$), 
Eq.~(\ref{eqn:en-thol}) is in complete agreement with Eq.~(3) of
Ref.~\cite{arguelles07}, providing an independent check of the algebra.
We note that, unlike the usual Bose-Hubbard model where $t_\sigma$ scales as
$1/d$ when $d \to \infty$, here $t_\sigma$ must scale as $1/\sqrt{d}$ 
when $d \to \infty$.

Given the two-particle and two-hole excitation energies, the mean-field phase
boundary for the Mott insulator--paired superfluid transition is determined by
(see Ref.~\cite{trefzger09} for a similar calculation)
\begin{align}
\label{eqn:mfe}
 1 = \frac{c_{\rm p}}{E_{\rm p} + c_{\rm p}} + \frac{c_{\rm h}}{E_{\rm h} + c_{\rm h}},
\end{align}
where 
$
c_{\rm p} = - 2(n_\uparrow+1)(n_\downarrow+1)z t_\uparrow t_\downarrow/U_{\uparrow\downarrow}
$
and
$
c_{\rm h} = - 2n_\uparrow n_\downarrow z t_\uparrow t_\downarrow/U_{\uparrow\downarrow}.
$
We note that, in the limit when $t_\uparrow = t_\downarrow = J$, 
$U_{\uparrow\uparrow} = U_{\downarrow\downarrow} = U$,
$U_{\uparrow\downarrow} = W \approx -U$, $n_\uparrow = n_\downarrow = m$, 
and $\mu_\uparrow = \mu_\downarrow = \mu$, Eq.~(\ref{eqn:mfe}) reduces to Eq.~(5) 
of Ref.~\cite{trefzger09} (after setting $U_{NN} = 0$ there). However, the terms 
that are proportional to $t_\uparrow t_\downarrow$ are not included in their 
definitions of the two-particle and two-hole excitation energies.
Solving Eq.~(\ref{eqn:mfe}) for $\mu_\uparrow + \mu_\downarrow$, we obtain
\begin{align}
\label{eqn:mfmu}
\mu_\uparrow + \mu_\downarrow = \frac{1}{2}\left[ a_{\rm p} - a_{\rm h} 
\pm \sqrt{(a_{\rm p} + a_{\rm h})^2 - 4c_{\rm p} c_{\rm h}} \right],
\end{align}
where $a_{\rm p} = E_{\rm p} + \mu_\uparrow + \mu_\downarrow$, 
$a_{\rm h} = E_{\rm h} - \mu_\uparrow - \mu_\downarrow$, and $\pm$ signs
correspond to the two-particle and two-hole branches, respectively.
Equation~(\ref{eqn:mfmu}) is the mean-field expression for 
the Mott insulator--paired superfluid transition boundary, and it is 
the main result of this paper.

\subsection{Typical phase diagrams}
\label{sec:pd}

In this section, we present typical phase diagrams in the 
$\mu_\uparrow + \mu_\downarrow$ versus $\sqrt{z} t_\uparrow$ plane, 
obtained directly from Eq.~(\ref{eqn:mfmu}). 
Similar to the usual Bose-Hubbard model, 
as hopping increases from zero, the range of the chemical potential about 
which the ground state is a Mott insulator decreases, and the Mott 
insulator phase disappears at a critical value of hopping, 
beyond which the system becomes a paired superfluid.

\begin{figure} [htb]
\centerline{\scalebox{0.6}{\includegraphics{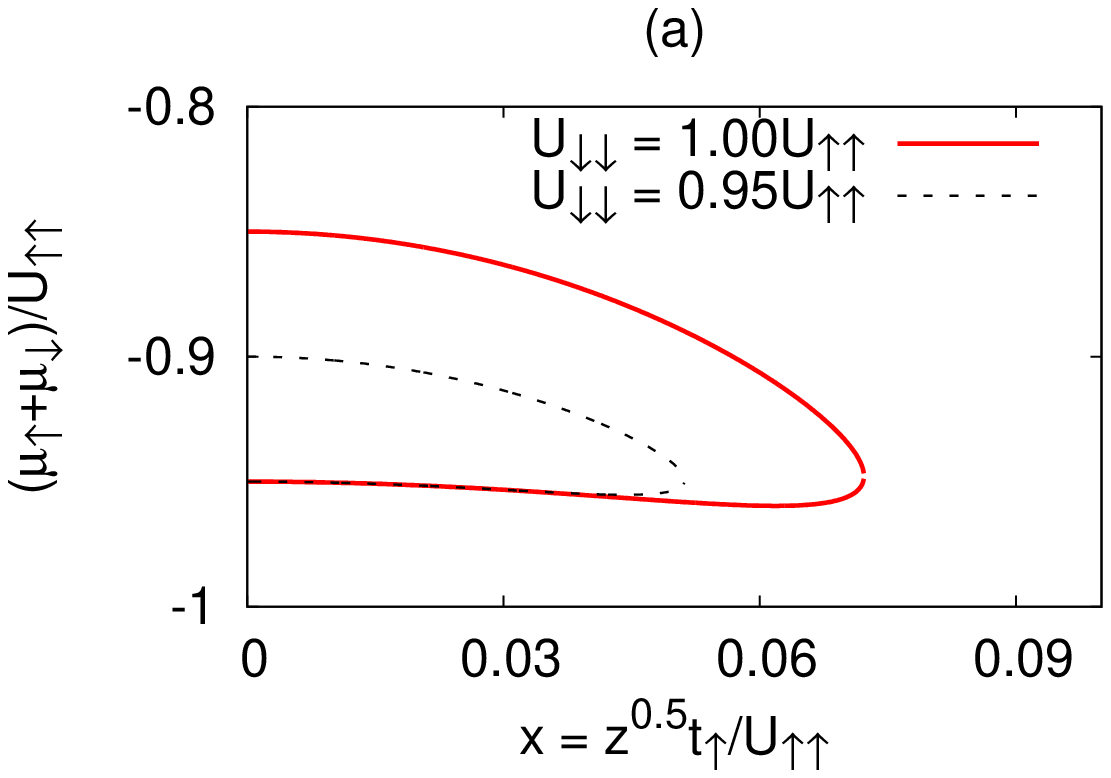}}}
\centerline{\scalebox{0.6}{\includegraphics{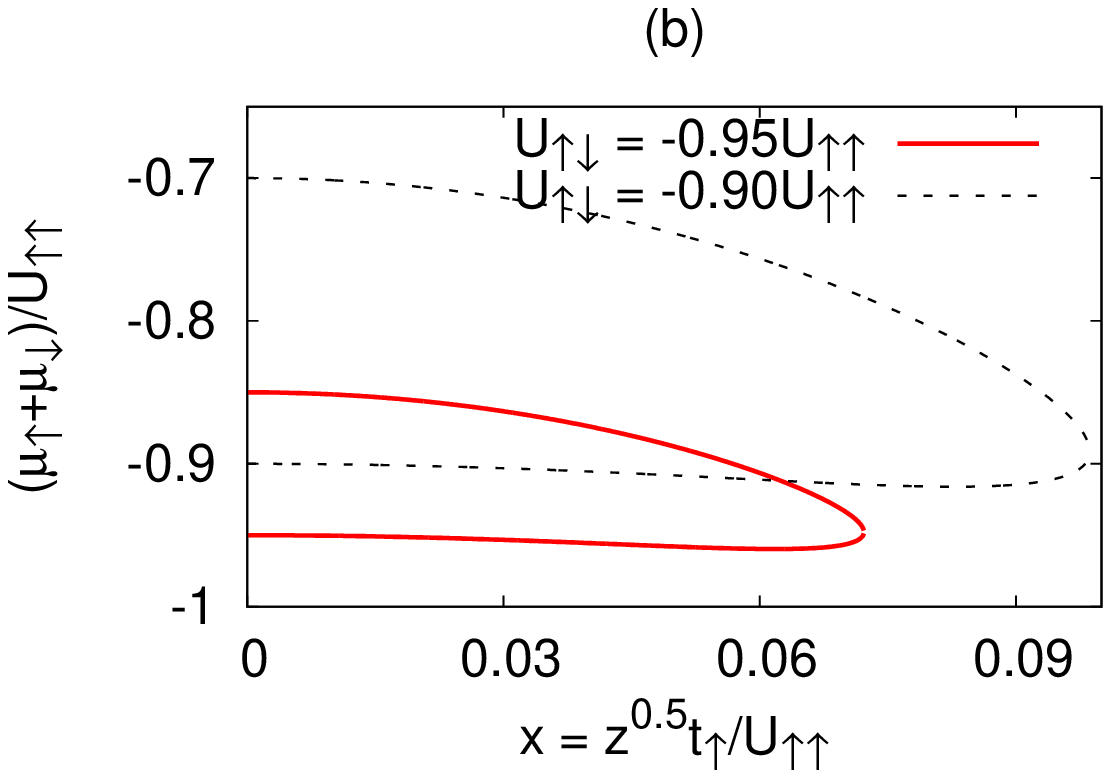}}}
\caption{\label{fig:1} (Color online)
The Mott insulator--paired superfluid phase transition boundaries are 
shown for the first lobe, i.e. $n_\uparrow = n_\downarrow = 1$, 
when $t_\uparrow = t_\downarrow$.
In (a) $U_{\uparrow\downarrow} = -0.95U_{\uparrow\uparrow}$, and 
in (b) $U_{\downarrow\downarrow} = U_{\uparrow\uparrow}$.
Note that the red curves correspond to the same data in both
of these figures.
}
\end{figure}

For instance, in Fig.~\ref{fig:1}, we show the Mott insulator--paired 
superfluid phase transition boundaries for the first lobe, 
i.e. $n_\uparrow = n_\downarrow = 1$, when $t_\uparrow = t_\downarrow$.
In Fig.~\ref{fig:1}(a), where we set $U_{\uparrow\downarrow} = -0.95U_{\uparrow\uparrow}$,
it is clearly seen that decreasing $U_{\downarrow\downarrow}$ favors the 
paired superfluid phase, as intuitively expected.
While in Fig.~\ref{fig:1}(b), where we set $U_{\downarrow\downarrow} = U_{\uparrow\uparrow}$,
it is clearly seen that decreasing the strength of $U_{\uparrow\downarrow}$ 
favors the Mott insulator phase (see the explanation below).
We also note a weak re-entrant quantum phase transition in both of the figures.

\begin{figure} [htb]
\centerline{\scalebox{0.6}{\includegraphics{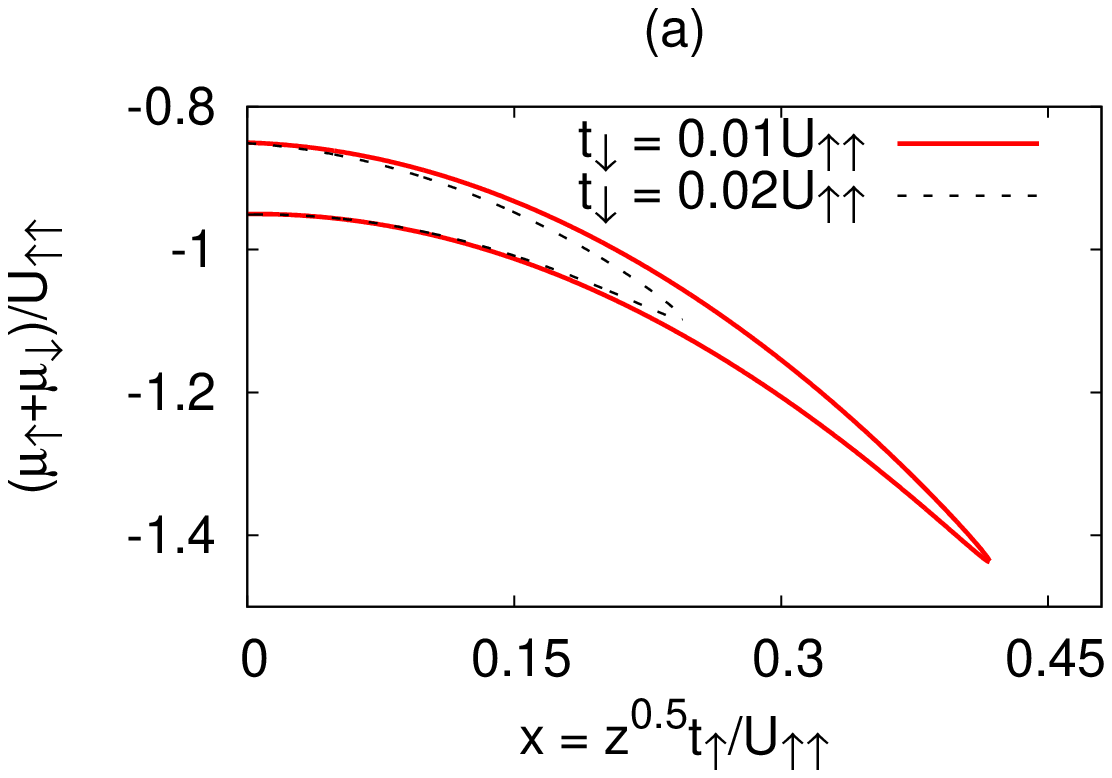}}}
\centerline{\scalebox{0.6}{\includegraphics{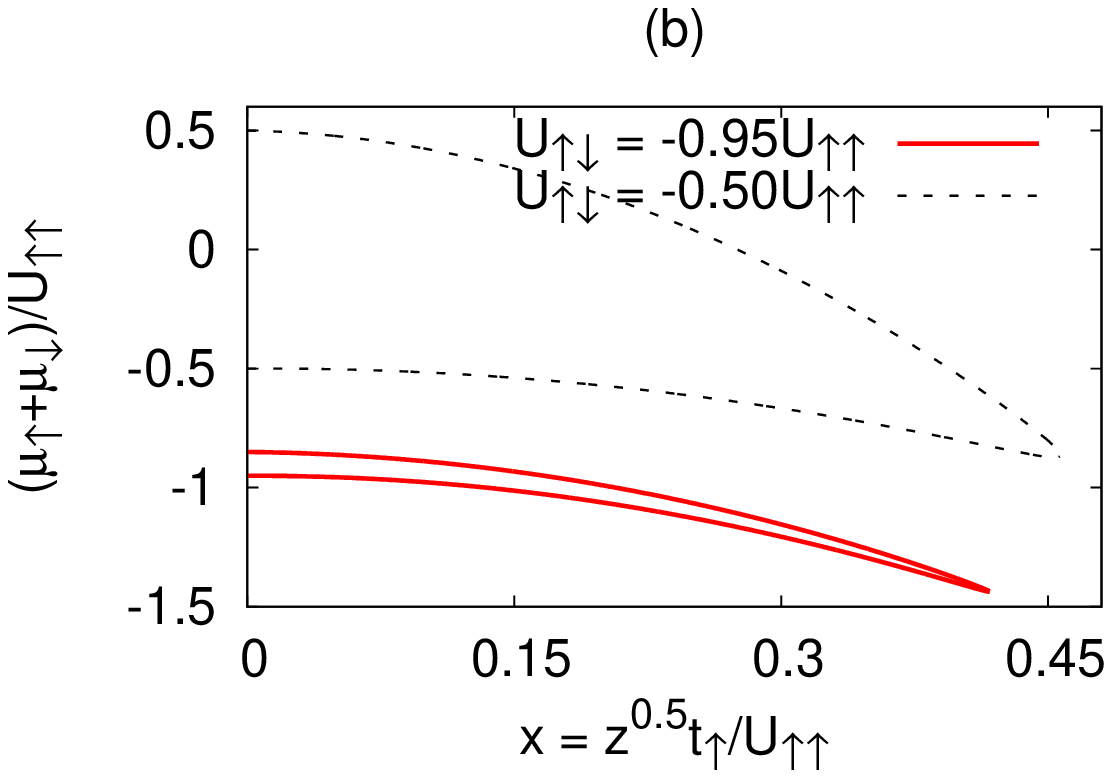}}}
\caption{\label{fig:2} (Color online)
The Mott insulator--paired superfluid phase transition boundaries are 
shown for the first lobe, i.e. $n_\uparrow = n_\downarrow = 1$, 
when $U_{\downarrow\downarrow} = U_{\uparrow\uparrow}$.
In (a) $U_{\uparrow\downarrow} = -0.95U_{\uparrow\uparrow}$, and 
in (b) $t_\downarrow = 0.01t_\downarrow$.
Note that the red curves correspond to the same data in both
of these figures.
}
\end{figure}

In addition, in Fig.~\ref{fig:2}, we show the Mott insulator--paired 
superfluid phase transition boundaries for the first lobe, 
i.e. $n_\uparrow = n_\downarrow = 1$, 
when $U_{\downarrow\downarrow} = U_{\uparrow\uparrow}$.
In Fig.~\ref{fig:2}(a), where we set $U_{\uparrow\downarrow} = -0.95U_{\uparrow\uparrow}$,
it is clearly seen that increasing $t_\downarrow$ favors the paired 
superfluid phase, as intuitively expected.
While in Fig.~\ref{fig:2}(b), where we set $t_\downarrow = 0.01t_\downarrow$,
it is clearly seen that decreasing the strength of $U_{\uparrow\downarrow}$ 
again favors the Mott insulator phase. However, compared to Figs.~\ref{fig:1}(a)
and~\ref{fig:1}(b), we note that the re-entrant quantum phase transition 
is much stronger in these figures.

Our results are consistent with the expectation that, 
for small $U_{\uparrow\downarrow}$, the location of the Mott insulator tip 
increases as a function of $U_{\uparrow\downarrow}$, because the presence 
of a nonzero $U_{\uparrow\downarrow}$ is what allowed this state to form in 
the first place. However, when the strength of $U_{\uparrow\downarrow}$ 
is larger than some critical value 
(approximately $\sqrt{U_{\uparrow\uparrow} U_{\downarrow\downarrow}}/2$), 
the location of the tip decreases, and it eventually vanishes exactly when 
$U_{\uparrow\downarrow}^2 = U_{\uparrow\uparrow} U_{\downarrow\downarrow}$, 
which may indicate an instability toward a collapse.
In addition, from Eq.~\ref{eqn:en-thol}, we expect a re-entrant quantum 
phase transition when
$
-(2n_\uparrow n_\downarrow/U_{\uparrow\downarrow}) z t_\uparrow t_\downarrow
-\sum_\sigma[ n_\sigma^2/U_{\uparrow\downarrow} 
	- (n_\sigma^2-1)/(2U_{\sigma\sigma}+U_{\uparrow\downarrow}) 
	+ 2n_\sigma(n_\sigma+1)/U_{\sigma\sigma}] z t_\sigma^2 < 0,
$
which occurs beyond a critical $U_{\uparrow\downarrow}$. When this expression 
is negative, its value is most negative for the first Mott lobe 
(i.e. $n_\uparrow = n_\downarrow = 1$), and therefore the effect is 
strongest there. However, its value increases and eventually becomes 
positive as a function of filling, and thus the re-entrant behavior becomes 
weaker as filling increases, and it eventually disappears 
beyond a critical filling.

\section{Conclusions}
\label{sec:conclusions}

In this paper, by restricting the two-species Bose-Hubbard Hamiltonian to the 
subspace of paired particles, and using perturbation theory, we derived an 
analytic mean-field expression for the Mott insulator--paired superfluid 
transition boundary. We found that, for small $U_{\uparrow\downarrow}$, 
the location of the Mott insulator tip increases as a function of 
$U_{\uparrow\downarrow}$, because the presence of a nonzero 
$U_{\uparrow\downarrow}$ is what allowed this state to form in 
the first place. However, when the strength of $U_{\uparrow\downarrow}$ 
is larger than some critical value 
(approximately $\sqrt{U_{\uparrow\uparrow} U_{\downarrow\downarrow}}/2$), 
the location of the tip decreases, and it eventually vanishes exactly when 
$U_{\uparrow\downarrow}^2 = U_{\uparrow\uparrow} U_{\downarrow\downarrow}$, 
which may indicate an instability toward a collapse.
Given that the interspecies interaction can be fine tuned in ongoing 
experiments, e.g. $^{41}$K-$^{87}$Rb~\cite{catani08, thalhammer08}
or homonuclear~\cite{gadway10} mixtures, via using Feshbach 
resonances, we hope that our predictions could be tested with 
ultracold atomic systems.

\section{Acknowledgments}
\label{sec:ack}

The author thanks Christian Trefzger for correspondence. This work 
is financially supported by the Marie Curie International Reintegration 
Grants (FP7-PEOPLE-IRG-2010-268239) and Scientific and Technological Research 
Council of Turkey (T\"{U}B$\dot{\mathrm{I}}$TAK).

\end{document}